\documentclass[10pt,pra,amssymb,nofootinbib,showpacs,twocolumn]{revtex4}
\usepackage{bm}
\usepackage{graphicx}
\usepackage{amsmath}
\newcommand{\bd}{\begin{displaymath}}
\newcommand{\ed}{\end{displaymath}}
\newcommand{\be}{\begin{equation}}
\newcommand{\ee}{\end{equation}}

\DeclareMathOperator{\Tr}{Tr} \makeatletter
\def\ud{\@ifnextchar[{\du}{\uu}}
\def\uu{\mathop{\mathrm{\mathstrut d\!}}\nolimits}
\def\du[#1]{\mathop{\mathrm{\mathstrut d}}\nolimits ^{#1}\!}
\makeatother

\begin{document}
\title{\bf On the Energy Increase in Space-Collapse Models}
\author{Angelo Bassi}
\email{bassi@mathematik.uni-muenchen.de} \affiliation{Mathematisches
Institut der Universit\"at M\"unchen, Theresienstr. 39, 80333
M\"unchen, Germany, \\The Abdus Salam International Centre for
Theoretical Physics, Strada Costiera 11, 34014 Trieste, Italy.}
\author{Emiliano Ippoliti}
\email{ippoliti@ts.infn.it} \affiliation{Department of Theoretical
Physics, University of Trieste, Strada Costiera 11, 34014 Trieste,
Italy \\ Istituto Nazionale di Fisica Nucleare, sezione di Trieste,
Via Valerio 2, 34127 Trieste, Italy}
\author{Bassano \surname{Vacchini}}
\email{bassano.vacchini@mi.infn.it}
\affiliation{Dipartimento di Fisica
dell'Universit\`a di Milano, Via Celoria 16, 20133 Milan, Italy \\
Istituto Nazionale di Fisica Nucleare, sezione di Milano,
Via Celoria 16, 20133 Milan, Italy}
\begin{abstract}
A typical feature of spontaneous collapse models which aim at
localizing wavefunctions in space is the violation of the principle
of energy conservation. In the models proposed in the literature the
stochastic field which is responsible for the localization mechanism
causes the momentum to behave like a Brownian motion, whose larger
and larger fluctuations show up as a steady increase of the energy
of the system. In spite of the fact that, in all situations, such an
increase is small and practically undetectable, it is an undesirable
feature that the energy of physical systems is not conserved but
increases constantly in time, diverging for $t \rightarrow \infty$.
In this paper we show that this property of collapse models can be
modified: we propose a model of spontaneous wavefunction collapse
sharing all most important features of usual models but such that
the energy of isolated systems reaches an asymptotic finite value
instead of increasing with a steady rate.
\end{abstract}
\pacs{03.65.Ta, 02.50.Ey, 05.40.--a} \maketitle

\section{Introduction}
\label{sec:introduction}

As it is well--known, space--collapse
models~\cite{drm1,drm2,drm6,bassi-drm,red1,drm3,ad1,{AdlerJPA01},ad3,pen1,dow1,drm4,
drm5,HalliwellPRD95} aim at a solution of the macro--objectification
or measurement problem in quantum mechanics by suitably modifying
the Schr\"odinger equation with non--linear stochastic terms. One of
the characteristic features of these models is the violation of
energy--conservation for isolated systems; such a violation is
determined by the stochastic process responsible for the
localization mechanism, which induces larger and larger fluctuations
of the wavefunction in the momentum space \cite{bassi-drm}: these
increasing fluctuations, in turn, determine the increase of the
energy of the system \cite{Ballentine-Gallis}. For typical values of
the parameters, such an increase is very small and undetectable with
present--day technology \cite{drm1}; still, one would wish to
restore the principle of energy conservation within space--collapse
models.

In this paper we make one step towards this direction: we analyze a
model of wavefunction space--collapse for which the energy of
isolated systems does not increase indefinitely, but reaches an
asymptotic finite value. An analogy with the quantum Brownian motion
will show that the stochastic process acts like a dissipative medium
which thermalizes the system to a fixed temperature (the temperature
of the medium) and will suggest how to restore perfect energy
conservation.

The paper is organized as follows: after a brief review of the main
features of dynamical reduction models (Sec.~\ref{sec:intro}), we
introduce the collapse--model which is the subject of the paper
(Sec.~\ref{sec:model}). In Sec.~\ref{sec:mast-equat-stat} we study
the master equation for the statistical operator originating from
the stochastic dynamics: this will provide the rationale for the
choice of the localization operator which defines the model. In
Secs.~\ref{sec:1gauss} to~\ref{sec:nrg} we will study in detail the
most relevant properties of the model: we will analyze the time
evolution of Gaussian wavefunctions (Sec.~\ref{sec:1gauss}); the
collapse mechanism and collapse probability
(Sec.~\ref{sec:asym-beh}); we will see that the physical predictions
of the model agree with very high accuracy with standard quantum
mechanical predictions and, at the same time, the model reproduces
classical mechanics at the macroscopic level
(Sec.~\ref{sec:micro-macro}). We will finally discuss the issue
related to energy non conservation (Sec.~\ref{sec:nrg}) and conclude
with some final remarks (Sec.~\ref{concl}).

\section{Structure of dynamical reduction models}
\label{sec:intro}

The typical structure of the evolution equation of collapse models
is\footnote{Of course, this form can be generalized, e.g. by adding
a finite (or countable) number of operators $A_{i}$, each of which
is coupled to a Wiener process $W_{i}$. Moreover, the Wiener
processes may be complex instead of real, as assumed here.}:
\begin{eqnarray} \label{nle}
d\,\psi_{t} & = &\left[ -\frac{i}{\hbar}\, H\, dt +
\sqrt{\lambda}\, \left( A - r_{t}\right) dW_{t} \nonumber\right. \\
& & - \left. \frac{\lambda}{2}\, \Big( A^\dag A - 2 r_{t} A +
r_{t}^2 \Big) dt\right] \psi_{t}, \quad
\end{eqnarray}
with:
\begin{equation} \label{efr}
r_{t} \; = \; \frac{1}{2}\,\langle\psi_t|( A +
A^\dag)|\psi_t\rangle.
\end{equation}
The operator $H$ is related to the standard quantum Hamiltonian of
the system, while $A$ is the {\it reduction operator}, i.e., the
operator on whose eigenmanifolds one wants to reduce the
statevector, as a consequence of the collapse mechanism; the
positive constant $\lambda$ sets the strength of the collapse
mechanism. The stochastic dynamics is governed by a standard Wiener
process $W_t$ defined on a probability space $(\Omega, {\mathcal F},
{\mathbb P})$. Note that the equation is non--linear but preserves
the norm of the statevector.

In the literature on collapse models, the operator $A$ is usually
assumed to be self--adjoint; in such a case, and if one further
assumes that it has only a discrete spectrum, it can be
proven~\cite{drm2} that the form of the second and third term of
Eq.~\eqref{nle}, which modify the standard Schr\"odinger evolution,
is such that:

\noindent 1. The statevector collapses with respect to the
``preferred basis'' generated by the operator $A$, i.e., for almost
all realizations of the stochastic process there exists an
eigenstate $|a_{n}\rangle$ of $A$ (depending of course on the
realization of the stochastic process) such that:
\begin{equation} \label{variance}
|\psi_{t}\rangle \; \longrightarrow \; |a_{n}\rangle \qquad\quad
\makebox{for $t \rightarrow \infty$}.
\end{equation}

\noindent 2. The average ${\mathbb E} [\langle P \rangle_{t}] \equiv
{\mathbb E} [ \langle \psi_{t} | P | \psi_{t} \rangle ]$ of any
operator $P$ which commutes with $A$ is constant in time, i.e., ${\mathbb E}[\langle \psi_{t} | P |
\psi_{t} \rangle ] = \langle \psi_{0} | P | \psi_{0} \rangle$. In
particular, if $P$ is a projection operator relative to an
eigenmanifold of $A$, this
together with property~\eqref{variance} implies that the
probability for the statevector to be reduced into the eigenmanifold
associated to $P$ is equal to $\langle \psi_{0} | P | \psi_{0}
\rangle$, i.e., to the probability that standard quantum mechanics
associates to the collapse, as a result of a measurement of the
operator $A$. This is due to the fact that $\langle \psi_{t} | P |
\psi_{t} \rangle$ turns out to be a martingale, thanks to the
particular structure of Eq.\eqref{nle}, so that by the martingale
property ${\mathbb E} [ \langle \psi_{t} | P | \psi_{t} \rangle
]=\langle \psi_{0} | P | \psi_{0} \rangle$ \cite{AdlerJPA01}.

It is important to keep in mind that the above results are valid
only when the standard Hamiltonian $H$ either commutes with $A$ or
is equal to zero; in all other cases, such results are only
approximate, the approximation depending on the value of $\lambda$.

\section{The model}
\label{sec:model}

In the literature, $A$ has been mainly taken equal to the position
operator $q$, or a function of $q$ like in the continuous version
\cite{drm2} of the original GRW model \cite{drm1}, the reason
being that the operator $q$ is the most natural candidate for
localizing wavefunctions in space. As anticipated in the previous
section, one consequence of such a choice is that the energy of
the system increases in time, diverging for time going to
infinity; it is then natural to wonder whether a different choice
for $A$ can preserve all most important features of collapse
models, but at the same time cure this energy non--conservation.
This problem finds a partly positive solution by making the
following Ansatz\footnote{A localization operator involving $q$
and $p$ has also been considered in ref. \cite{bgrw} (note however
that the form of the localization operator is different from ours)
but with a different aim, i.e., that of studying whether the
presence of a $p$-term instead of only a $q$-term can improve the
localization mechanism. The authors prove that, for any physically
reasonable choice of the parameters of their model, such term does
not affect in an appreciable way the collapse mechanism. Here we
show that a $p$-term is important as it can change the time
evolution of the mean energy, avoiding it to increase constantly
in time. The authors of ref. \cite{HalliwellPRD95} analyze a
stochastic differential equation similar to our Eq.~\eqref{nle}
where both a $q$ and a $p$ term are present: they mainly focus
their attention on the application of the formalism to the theory
of open quantum systems and decoherent histories. One of their
main result is a localization theorem which we will apply to our
model to prove the collapse of wavefunctions to localized states.}
for $A$:
\begin{equation} \label{efa}
A \; = \; q \, + \, i\,\frac{\alpha}{\hbar}\, p,
\end{equation}
where $p$ is the momentum operator. Moreover, we define the operator
$H$ as follows:
\begin{equation}
H \; = \; H_{0} \, + \, \frac{\lambda\alpha}{2}\, \{ q, p \},
\end{equation}
where $H_{0}$ is the standard quantum Hamiltonian. In the following
sections we will analyze the most relevant physical properties of
the model and we will focus our attention to the case of a {\it
free} particle: $H_{0} = p^2/2m$, where $m$ is the mass of the
particle.

The model is defined in terms of the two constants $\lambda$ and
$\alpha$; for reasons which will be clear in the following, we will
assume them to vary with the mass of the particle as follows:
\begin{equation} \label{eqp}
\lambda \; = \; \frac{m}{m_{0}}\, \lambda_{0} \qquad\qquad \alpha
\; = \; \frac{m_{0}}{m}\, \alpha_{0},
\end{equation}
where $m_0$ is a reference mass which we choose to be equal to
that of a nucleon while $\lambda_0$ and $\alpha_0$ are fixed
constants which we take equal to:
\begin{eqnarray}
\lambda_{0} & \simeq & 10^{-2} \; \makebox{m$^{-2}$ sec$^{-1}$}, \label{eql}\\
\alpha_{0} & \simeq & 10^{-18} \; \makebox{m$^{2}$}. \label{eqa}
\end{eqnarray}
As it will be shown in Sec.~\ref{sec:micro-macro} this numerical
choice for the parameters guarantees that the model reproduces
almost exactly the physical predictions of standard quantum
mechanics at the microscopic level and reproduces classical
mechanics at the macroscopic level.

Before concluding this section, we note that, in order to find the
solutions of Eq. (\ref{nle}) and to study their properties, it is
convenient to consider also a linearized version of Eq.~\eqref{nle}
\cite{drm2,drm6}:
\begin{equation} \label{le}
d\,\phi_{t}(x) = \left[ -\frac{i}{\hbar} H \, dt \, + \,
\sqrt{\lambda}\, A \, d\xi_t \, - \, \frac{\lambda}{2}\, A^\dag
A\, dt \right] \phi_{t}(x), \quad
\end{equation}
where $\xi_{t}$ is a standard Wiener processes defined on a new
probability space $(\Omega, {\mathcal F}, {\mathbb Q})$. In ref.
\cite{bassi-drm} the relation between the probability measures
${\mathbb Q}$ and ${\mathbb P}$ and the relation between the
stochastic processes $W_{t}$ and $\xi_{t}$ are discussed. Here we
simply recall how one can use the above linear equation to find a
solution of Eq. (\ref{nle}):
\begin{enumerate}
\item Find the solution $\phi_{t}$ of Eq. (\ref{le}), with the
initial condition $\phi_{0} = \psi_{0}$.

\item Normalize the solution\footnote{If $\| \phi_{t} \| = 0$, then
one can set $\psi_{t}$ equal to any fixed unitary vector.}:
$\phi_{t} \rightarrow \psi_{t} = \phi_{t}/\| \phi_{t} \|$.

\item Make the substitution: $d\xi_{t} \rightarrow d W_{t} =
d\xi_{t} - 2\sqrt{\lambda} r_{t}$;
\end{enumerate}
the wavefunction $\psi_{t}$ thus obtained is a solution of Eq.
(\ref{nle}).

\section{The master-equation for the statistical operator}
\label{sec:mast-equat-stat}

In order to better understand the modifications to the
Schr\"odinger dynamics induced by Eq.~\eqref{nle} and the
motivations for the precise choice of its form, apart from the
martingale structure, and in particular in order to see why the
choice (\ref{efa}) for $A$ can partially cure the problem of the
energy increase, it is worthwhile considering the related equation
for the statistical operator $\rho_{t} \equiv {\mathbb E}[
|\psi_{t} \rangle\langle \psi_{t} |]$, which is given by
\begin{eqnarray} \label{eq:1}
\frac{d}{dt}\, \rho_t &= & - \frac{i}{\hbar}\, \left[
H_0, \rho_t \right] \, - \frac{\lambda}{2}\, \left[q, \left[ q,
\rho_t \right]
\right] \nonumber \\
& & - \frac{\lambda \alpha^2}{2\hbar^2}\, \left[p, \left[ p, \rho_t
\right] \right]\, - \, i\, \frac{\lambda \alpha}{\hbar}\, \left[ q,
\left\{ p, \rho_t \right\} \right],
\end{eqnarray}
that is the typical structure of master-equation for the quantum
description of Brownian motion, where both friction and diffusion
are taken into account and positivity of the statistical operator is
granted at all times.
The obvious difference between Eq.~\eqref{eq:1} and the master-equation
for quantum Brownian motion lies in the meaning of the coefficients,
here related to the two fundamental constants of the model $\lambda$
and $\alpha$, rather than to the friction coefficient and the
temperature of the bath. The quantum Brownian motion master-equation
is in fact given by~\cite{art3,art6,art10}
\begin{eqnarray}
\label{eq:1bis} \frac{d}{dt}\, \rho_t &= & - \frac{i}{\hbar}\,
\left[ H_0, \rho_t \right] \, - \gamma\frac{2M}{\beta\hbar^2}\,
\left[q, \left[ q, \rho_t \right]
\right] \nonumber \\
& & - \gamma\frac{\beta}{8M}\, \left[p, \left[ p, \rho_t \right]
\right]\, - \,\frac{i}{\hbar}\gamma \, \left[ q, \left\{ p, \rho_t
\right\} \right],
\end{eqnarray}
where $\gamma$ is the friction coefficient and $\beta$ the inverse
temperature of the background medium; the second and third term at
r.h.s. account for diffusion, 
with coefficients proportional to the
squared thermal wavelength ${\Delta x}^2_{\rm \scriptscriptstyle
  th}={\beta\hbar^2}/{4M}$ and the squared thermal momentum spread ${\Delta p}^2_{\rm \scriptscriptstyle
  th}={M}/{\beta}$ satisfying the minimum uncertainty
relation ${\Delta p}_{\rm \scriptscriptstyle th}{\Delta x}_{\rm
  \scriptscriptstyle th}={\hbar}/{2}$,
while the last is due to friction and
ensures that the energy of the test particle asymptotically goes to
the equipartition value depending only on the temperature of the
bath. Note that in the quantum description friction, which
accounts for the finite energy growth, is of necessity related to
diffusion in order to preserve the Heisenberg uncertainty relation~\cite{LindbladQBM,art3,art7}.
A fundamental result, in order to understand how Eq.~\eqref{eq:1}
and therefore the striking similarity with quantum Brownian motion
appears, is Holevo's characterization of translation-covariant
generators of quantum-dynamical
semigroups~\cite{HolevoJMP-HolevoRMP32-HolevoRMP33-HolevoRAN},
according to which further important restrictions can be put on the
operators appearing in the general so-called Lindblad structure,
once symmetry under translations is taken into account.  In fact,
according to Holevo's result, if the generator of the dynamics
$\mathcal{L}$ is translation-covariant, i.e., commutes with the
action of the unitary representation of translations $U (a)
=\exp[{-(i/\hbar)ap}]$
\begin{equation}
   \label{eq:2}
   \mathcal{L}[U(a)\rho U^{\scriptscriptstyle \dagger}(a)]=
U (a)\mathcal{L}[\rho]U^{\scriptscriptstyle \dagger}(a)
\end{equation}
for all real $a$, then its general structure, given that $q$ appears
at most quadratically, is the following
\begin{equation}
   \label{eq:3}
    \mathcal{L}[\rho]=-\frac{i}{\hbar}[H
   (p),\rho]
+\mathcal{L}_{G}[\rho],
\end{equation}
with $H (p)$ a self-adjoint operator only depending on the momentum
operator $p$ and  $\mathcal{L}_{G}$ (where $G$ stands for Gaussian)
is given by
\begin{align}
   \label{eq:4}
   \mathcal{L}_{G}[\rho]
=&-{i \over \hbar}
        \left[a_0 q+
        H_{\mathrm{\scriptscriptstyle eff}} (q,p)
        ,\rho
        \right]
\\ \nonumber
&+ \left[K \rho K ^{\dagger} -\frac{1}{2}\left\{K ^{\dagger}K
,\rho\right\} \right],
\end{align}
with
\begin{align*}
K  &=a_1 q +L_1  (p)  \qquad a_0, a_1\in \mathbb{R}
\\
H_{\mathrm{\scriptscriptstyle eff}} (q,p)&=\frac{\hbar}{2i} a_1 (q
L_1  (p) -L_1 ^{\dagger} (p)q )
\end{align*}
and $L_1(p)$ a generally complex function of its argument. The
requirement of translational invariance is a natural and compelling
one for dynamical reduction models, since the modification of
quantum mechanics by a universal noise should by no way break the
homogeneity of space, introducing some preferred space location. The
restriction to mappings at most quadratic in the position operator
$q$ is linked to the fact that we are looking for a generalization
of the most simple dynamical reduction model where $A=q$ and the
associated master-equation is given by
\begin{eqnarray}
\label{eq:5} \frac{d}{dt}\, \rho_t & =& - \frac{i}{\hbar}\, \left[
H_0, \rho_t \right] \, - \frac{\lambda}{2}\, \left[q, \left[ q,
\rho_t \right] \right] ,
\end{eqnarray}
often considered in the literature (see e.g. \cite{bassi-drm} and
references therein) even though leading to a steady energy
increase. In view of Eq.~\eqref{eq:4} the most straightforward
extension of Eq.~\eqref{eq:5} including a friction term
proportional to velocity is obtained setting $a_0=0$, i.e., no
external constant force since we are considering the modification
to Schr\"odinger dynamics for a free particle,
$a_1=\sqrt{\lambda}$ and $L_1 (p) =
i\sqrt{\lambda}(\alpha/\hbar)p$, thus directly obtaining the
Ansatz given in Eq.~\eqref{efa}. With this choice of functions and
parameters Eq.~\eqref{eq:3} gives
\begin{eqnarray}
\label{eq:6} \frac{d}{dt}\, \rho_t &=& - \frac{i}{\hbar}\, \left[
H_0+\frac{\lambda
     \alpha}{2}\, \left\{ q, p \right\}, \rho_t \right]
\nonumber \\
&&+\lambda \left[A \rho A^{\dagger} -\frac{1}{2}\left\{A^{\dagger}A
,\rho\right\} \right]
\end{eqnarray}
with $A  =  q + i\,(\alpha/\hbar)\, p$, as in Eq.~\eqref{efa}, which
is immediately seen to be equivalent to Eq.~\eqref{eq:1}, thus
giving the rationale for our choice for the operator $A$.

Note that looking at Eq.~\eqref{eq:6} one might erroneously be led
to think that the modification to Schr\"odinger dynamics amounts
to a change in the Hamiltonian plus a correction due to a
universal noise given by a mapping in the Lindblad form with a
single so-called Lindblad operator $A$. This standpoint,
implicitly adopted in~\cite{HalliwellPRD95}, and which has often
led to confusion~\cite{commentiME}, is actually misleading. The result by Lindblad, which is strictly
speaking only valid when the generator is bounded, but as shown by
Lindblad himself~\cite{LindbladQBM} and by the quoted results of
Holevo also holds for the case at hand, states that the generator
of a completely positive quantum-dynamical semigroup has the form
\begin{eqnarray}
\label{eq:7} \frac{d}{dt}\, \rho_t &=& - \frac{i}{\hbar}\, \left[
H, \rho_t \right]
\nonumber \\
&&+\sum_i \left[L_i\rho L_i^{\dagger}
-\frac{1}{2}\left\{L_i^{\dagger}L_i,\rho\right\} \right],
\end{eqnarray}
where the self-adjoint operator $H$ is not necessarily the free
Hamiltonian of the system, giving its dynamics when it is not
coupled to the environment or some noise source. On the contrary it
usually happens, e.g. when the Lindblad structure appears in the
reduced description of a system coupled to some reservoir, that in the
commutator at r.h.s. of~\eqref{eq:7} an operator appears which is
the sum of the free Hamiltonian and some other self-adjoint
operator, this other contribution disappearing together with the
rest of the Lindblad form when the coupling vanishes, as it
correctly happens in Eq.~\eqref{eq:6} if the fundamental constant
$\lambda$ is set to zero. The general structure~\eqref{eq:7} cannot
be thought of as being made up of two distinct parts, since the
Lindblad characterization pertains to the structure as a whole. Note
however that the free Hamiltonian can still be put into evidence
in~\eqref{nle} according to
\begin{multline} \label{nle:bis}
d\,\psi_{t} = \left[ -\frac{i}{\hbar}\, H_0\, dt +
\sqrt{\lambda}\, \left( A - r_{t}\right) dW_{t} \right. \\
 - \left. \frac{\lambda}{2}\, \Big( A^\dag A - 2 r_{t} A +
    r_{t}^2+\frac{1}{2} (A^2 -A^\dag{^2}) \Big) dt\right] \psi_{t}, \quad
\end{multline}
even though in this equivalent expression the martingale structure
is less evident.

\section{Single--Gaussian solution}
\label{sec:1gauss}

Gaussian wavefunctions are very special and they are often used to
represent typical physical situations; we now show that, as for the
standard Schr\"odinger equation, our stochastic equation preserves
the form of Gaussian wavefunctions and, at the same time, we analyze
their evolution in time. Let us then consider the following class of
wavefunctions:
\begin{equation}
\label{gsol}
\phi_{t}(x) = \makebox{exp}\left[ - a_{t} (x - \overline{x}_{t})^2
+ i \overline{k}_{t}x + \gamma_{t}\right],
\end{equation}
where $a_{t}$ and $\gamma_{t}$ are complex functions of time, while
$\overline{x}_{t}$ and $\overline{k}_{t}$ are real. By following the
procedure outlined in ref. \cite{bassi-drm}, one can show that the
above parameters obey the following stochastic differential
equations\footnote{The superscripts ``R'' and ``I'' denote,
respectively, the real and imaginary parts of the corresponding
quantities.}:
\begin{eqnarray}
da_{t} & = & \left[ - \frac{2i\hbar}{m}\, a_{t}^2
-4\lambda\alpha\,a_t + \lambda\right]
dt \label{pp1},\\
d \overline{x}_{t} & = & \frac{\hbar}{m}\overline{k}_{t}\,dt +
\sqrt{\lambda} \left[ \frac{1}{2a_{t}^{\makebox{\tiny R}}} -
\alpha \right] dW_{t}, \label{pp2} \\
d \overline{k}_{t} & = & -2\lambda \alpha\,\overline{k}_{t}\, dt -
\sqrt{\lambda}\, \frac{a_{t}^{\makebox{\tiny
I}}}{a_{t}^{\makebox{\tiny R}}} dW_{t}. \label{pp3}
\end{eqnarray}
We have omitted the equation for $\gamma_t$ since the real part can
be absorbed into the normalization factor, while the imaginary part
represents an irrelevant global phase.

\subsection{The time evolution of $a_{t}$}

The parameter $a_{t}$ is particularly important since it is related
to the spread of the wavefunction (\ref{gsol}) in position and
momentum, according to the following expressions:
\begin{eqnarray}
      \label{spm}
\sigma_{q}(t) & \equiv & \sqrt{\langle q^2 \rangle - \langle q
\rangle^2 } \; = \; \frac{1}{2}\sqrt{\frac{1}{a_{t}^{\makebox{\tiny
R}}}}, \nonumber \\
\sigma_{p}(t) & \equiv & \sqrt{\langle p^2 \rangle - \langle p
\rangle^2 } \; = \; \hbar\,\sqrt{\frac{(a_{t}^{\makebox{\tiny R}})^2
+ (a_{t}^{\makebox{\tiny I}})^2}{a_{t}^{\makebox{\tiny R}}}},
\end{eqnarray}
Eq. (\ref{pp1}) for $a_{t}$ can be easily solved; one gets:
\begin{equation}
a_{t} \; = \; -\frac{1}{2} \left[ A + i B \tanh \left(
\frac{\hbar}{m}\, B\, t + k \right)\right],
\end{equation}
with:
\begin{equation}
A \; = \; -2i\frac{\lambda \alpha m}{\hbar}, \qquad B \; = \;
\sqrt{\frac{4\lambda^2 \alpha^2 m^2}{\hbar^2} + i\frac{2\lambda
m}{\hbar}};
\end{equation}
the constant $k$ sets the initial condition $a_{0}$.

After a tedious calculation, one can write explicitly the time
evolution of the real and imaginary parts\footnote{The two
parameters $\varphi_{1}$ and $\varphi_{2}$ are functions of the
initial conditions.} of $a_{t}$:
\begin{equation}
a_{t}^{\makebox{\tiny R}}  = \frac{m\omega}{2\sqrt{2} \hbar}\,
\frac{\sin\theta\sinh (\omega_1 t + \varphi_{1}) + \cos\theta\sin
(\omega_2 t + \varphi_{2})}{ \cosh (\omega_1 t + \varphi_{1}) + \cos
(\omega_2 t +\varphi_{2})}, \label{rpa}
\end{equation}
\begin{eqnarray}
a_{t}^{\makebox{\tiny I}} & = & \frac{-m\omega}{2\sqrt{2} \hbar}
\left[ \frac{\cos\theta\sinh (\omega_1 t + \varphi_{1}) -
\sin\theta\sin (\omega_2 t + \varphi_{2})}{ \cosh (\omega_1 t +
\varphi_{1}) + \cos (\omega_2 t +\varphi_{2})}\right. \nonumber
\\
& & - \left.
{\frac{2\sqrt{2}\lambda \alpha}{\omega}}
\right]\label{ipa}
\end{eqnarray}
where we have introduced the following two frequencies:
\begin{equation}
\omega_1 = \sqrt{2}\, \omega \, \cos \theta \qquad \omega_2 =
\sqrt{2}\, \omega \, \sin \theta,
\end{equation}
the frequency $\omega$ and the angle $\theta$ being defined as
follows:
\begin{eqnarray}
\omega & = & 2\,\sqrt[4]{4\lambda_{0}^4\alpha_{0}^4 \, + \,
\frac{\lambda_{0}^2 \hbar^2}{m_{0}^2}} \; \simeq \; 10^{-5} \;
\makebox{sec$^{-1}$}, \\
\theta & = &
\frac{1}{2}\tan^{-1}\left[\frac{\hbar}{2\lambda_{0}\alpha_{0}^2
m_{0}} \right] \; \simeq \; \frac{\pi}{4};
\end{eqnarray}
note that, due to the specific dependence of both $\lambda$ and
$\alpha$ on $m$ as given by Eq.~\eqref{eqp}, both $\omega$ and
$\theta$ are independent of the mass of the particle, and thus are
two constants of the model. Note also that
--- as it is easy to prove
--- if $a_{0}^{\makebox{\tiny R}} > 0$, then $a_{t}^{\makebox{\tiny
R}} > 0$ for any subsequent time $t$: this implies that Gaussian
solutions do not diverge in time.

\subsection{The spread in position and momentum}

According to (\ref{spm}), the spread in position of the Gaussian
wavefunction (\ref{gsol}) evolves in time as follows\footnote{Note
that the evolution of $\sigma_{q}(t)$ (and also of $\sigma_{p}(t)$)
is {\it deterministic} and depends on the noise $W_{t}$ only
indirectly, through the constant $\lambda$.}:
\begin{equation} \label{sx}
\sigma_{q}(t) = \sqrt{\frac{\hbar}{\sqrt{2}m \omega}\frac{\cosh
(\omega_1 t + \varphi_{1}) + \cos (\omega_2 t +
\varphi_{2})}{\sin\theta\sinh (\omega_{1} t + \varphi_{1}) +
\cos\theta\sin (\omega_{2} t + \varphi_{2})}}
\end{equation}
Here we can see one of the effects of the reduction mechanism:
while in the standard quantum case the spread (in position) of the
wavefunction of a free quantum particle increases in time,
diverging for $t \rightarrow \infty$, the spread according to our
model reaches the asymptotic {\it finite} value:
\begin{equation} \label{eq:as-spread-pos}
\overline{\sigma}_{q} \; \equiv \; \sigma_{q}(\infty) \; = \;
\sqrt{\frac{\hbar}{\sqrt{2}m \omega \sin \theta}} \; \simeq \;
\left( 10^{-15} \sqrt{\frac{\makebox{Kg}}{m}}\, \right)\,
\makebox{m};
\end{equation}
The asymptotic spread decreases for increasing values of the mass of
the particle, this property entailing that, as we shall discuss in
more detail in Sec.~\ref{sec:micro-macro}, wavefunctions of
macroscopic objects are almost always very well localized in space,
so well that they practically behave like point--like particles.

The time evolution for $\sigma_{p}(t)$ can also be obtained from
(\ref{spm}) and, as it happens for the spread in position, also the
spread in momentum asymptotically reaches a finite value, which is:
\begin{eqnarray} \label{eq:as-spread-mom}
\overline{\sigma}_{p} \; \equiv \; \sigma_{p}(\infty) & = &
\sqrt{\frac{\hbar m \omega}{2 \sqrt{2}}\;
\frac{\sin^2\theta + (\cos\theta-\kappa)^2}{\sin\theta}} \nonumber \\
& \simeq & \left( 10^{-19} \sqrt{\frac{m}{\makebox{Kg}}}\,
\right)\, \frac{\makebox{Kg m}}{\makebox{sec}}
\end{eqnarray}
where:
\begin{equation}
\kappa \; = \; \frac{2\sqrt{2} \lambda_{0} \alpha_{0}}{\omega} \;
\simeq \; 10^{-14}.
\end{equation}

To conclude the section, we compute the product of the two
asymptotic spreads:
\begin{equation} \label{pin}
\overline{\sigma}_{q} \cdot \overline{\sigma}_{p} \; = \;
\frac{\hbar}{2} \;
\sqrt{1+\frac{(\cos\theta-\kappa)^2}{\sin^2 \theta}}
,
\end{equation}
which is almost equal to $\hbar/\sqrt{2}$, with our choice
(\ref{eql}) and (\ref{eqa}) for the parameters. Note however that
for any choice of $\lambda_{0}$ and $\alpha_{0}$ Heisenberg
uncertainty relations are fulfilled.

In accordance with \cite{di1}, any Gaussian solution having this
asymptotic values for the spread in position and momentum will be
called a ``stationary solution'' of Eq. (\ref{nle}). Of course, the
term ``stationary'' does not mean that such wavefunctions do not
evolve in time; as a matter of fact (see the following discussion)
they always undergo a random motion both in position and momentum
which never stops. The term ``stationary'' refers only to the shape
of the wavefunction: stationary solutions are special wavefunction
which are {\it Gaussian} and with a {\it fixed} spread in position
and momentum, given by Eqs.~\eqref{eq:as-spread-pos}
and~\eqref{eq:as-spread-mom}.

\subsection{The mean in position and momentum}

The mean $\langle q \rangle_{t}$ in position of the wavefunction and
the mean $\langle p \rangle_{t}$ in momentum satisfy the following
stochastic differential equations which can be derived from
Eqs.~\eqref{pp2} and~\eqref{pp3}:
\begin{eqnarray}
d \langle q \rangle_{t} & = & \frac{1}{m}\,\langle p
\rangle_{t}\,dt + \sqrt{\lambda} \left[
\frac{1}{2a_{t}^{\makebox{\tiny R}}} -
\alpha \right] dW_{t}, \label{ep} \\
d \langle p \rangle_{t} & = & -2\lambda \alpha\,\langle p
\rangle_{t}\, dt - \sqrt{\lambda}\hbar\,
\frac{a_{t}^{\makebox{\tiny I}}}{a_{t}^{\makebox{\tiny R}}}
dW_{t}. \label{em}
\end{eqnarray}
Their average values evolve as follows:
\begin{eqnarray} \label{mpa}
m\frac{d}{dt}\,{\mathbb  E}\left[ \langle q \rangle_{t} \right] &
= & {\mathbb  E}\left[ \langle p \rangle_{t} \right], \\
\frac{d}{dt}\,{\mathbb  E}\left[ \langle p \rangle_{t} \right] & =
& -2\lambda\alpha\, {\mathbb E}\left[ \langle p \rangle_{t}
\right].
\end{eqnarray}
The first equation reproduces the classical relation between
position and momentum of a particle while the second equation
predicts that the momentum decays exponentially in time:
\begin{equation}
{\mathbb  E}\left[ \langle p \rangle_{t} \right] \; = \; \langle p
\rangle_{0}\, e^{\displaystyle -2\lambda\alpha t},
\end{equation}
with
\begin{equation} \label{etr}
2\lambda\alpha \; = \; 2\lambda_{0}\alpha_{0} \; \simeq \;
10^{-20} \; \makebox{sec$^{-1}$},
\end{equation}
which is an extremely slow decay rate, not depending on the mass of
the particle.

For completeness we consider also the covariance matrix
\begin{eqnarray}
C(t) & = & {\mathbb E} \left[ \left[
\begin{array}{c}
\langle q \rangle_{t} - {\mathbb E}[\langle q \rangle_{t}] \\
\langle p \rangle_{t} - {\mathbb E}[\langle p \rangle_{t}]
\end{array}
\right] \cdot \left[
\begin{array}{c}
\langle q \rangle_{t} - {\mathbb E}[\langle q \rangle_{t}] \\
\langle p \rangle_{t} - {\mathbb E}[\langle p \rangle_{t}]
\end{array}
\right]^{\top} \right] \nonumber \\
& & \equiv \left[
\begin{array}{cc}
C_{q^2}(t) & C_{qp}(t) \\
C_{pq}(t) & C_{p^2}(t)
\end{array}
\right], \nonumber
\end{eqnarray}
whose coefficients satisfy the following equations:
\begin{eqnarray}
\frac{d}{dt}\, C_{q^2}(t) & = & \frac{2}{m}\, C_{qp}(t) +
\lambda \left(\frac{1}{2 a_{t}^{\makebox{\tiny R}}}
- \alpha \right)^2 \label{eq1c} \\
\frac{d}{dt}\, C_{qp}(t) & = & \frac{1}{m}\, C_{p^2}(t) -
2\lambda\alpha C_{qp}(t) - \lambda \hbar
\frac{a_{t}^{\makebox{\tiny I}}}{a_{t}^{\makebox{\tiny
R}}}\left(\frac{1}{2
a_{t}^{\makebox{\tiny R}}} - \alpha \right)  \nonumber \\
& &\label{eq2c}  \\
\frac{d}{dt}\, C_{p^2}(t) & = & -4\lambda\alpha\, C_{p^2}(t) +
\lambda \hbar^2 \left( \frac{a_{t}^{\makebox{\tiny
I}}}{a_{t}^{\makebox{\tiny R}}} \right)^2. \label{eq3c}
\end{eqnarray}
In Sec. \ref{sec:macro} we will discuss the physical implications of
the above equations in connection with the dynamics of macroscopic
objects.

\section{Asymptotic behavior of the general solution}
\label{sec:asym-beh}

In the previous section we have seen that any Gaussian solution
converges towards a stationary solution i.e., towards a Gaussian
wavefunction with a fixed finite value both for the spread in
position and momentum, given by Eqs.~\eqref{eq:as-spread-pos}
and~\eqref{eq:as-spread-mom}. In this section we prove that the
spread $\sigma_{q}(t)$ of {\it any} wavefunction converges with
probability one towards $\overline{\sigma}_{q}$: this means that any
initial wavefunction converges to a localized solution; for the
proof we will follow the same strategy of Ref.
\cite{HalliwellPRD95}.

\subsection{The reduction process}

It is easy to see that a Gaussian stationary solution is an eigenstate of the
operator:
\begin{equation}
O \; = \; p \, - \, 2 i \hbar a_{\infty} q,
\end{equation}
where
\begin{equation}
a_{\infty} \; = \; \frac{m\omega}{2\sqrt{2}\hbar}\left[ \sin\theta -
i \left(\cos\theta - \kappa\right) \right];
\end{equation}
the proof basically consists in showing that the variance
\begin{equation} \label{var}
\sigma_{O}^2(t) \; \equiv \; \langle \psi_{t} |[ O^{\dagger} -
\langle O^{\dagger} \rangle] [O - \langle O \rangle] | \psi_{t}
\rangle
\end{equation}
of the operator $O$ vanishes for $t \rightarrow \infty$.

The first step is to re--write $\sigma_{O}^2(t)$ in terms of the
variances associated to the operators $q$ and $p$:
\begin{equation}
\label{AO2}
\sigma_{O}^2(t) \; = \; \sigma_{p}^2(t) +
\frac{\overline{\sigma}_{p}^2}{\overline{\sigma}_{q}^2}\,
\sigma_{q}^2(t) - 2\,
\frac{\overline{\sigma}_{q,p}^2}{\overline{\sigma}_{q}^{2}}\,
\sigma_{q,p}^2(t) - \frac{\hbar^2}{2 \overline{\sigma}_{q}^{2}},
\end{equation}
where we have defined:
\begin{eqnarray}
\sigma_{q,p}^2(t) & = & \frac{1}{2}\left[ \langle \psi_{t} |[ q -
\langle q \rangle] [p - \langle p \rangle] | \psi_{t} \rangle +
\right.
\nonumber \\
& & \;\;\;\; \left. \langle \psi_{t} |[ p - \langle p \rangle] [q -
\langle q \rangle] | \psi_{t} \rangle \right],
\end{eqnarray}
 so that for a Gaussian wavefunction such as~\eqref{gsol}
 \begin{equation}
 \label{sigmaxp}
   \sigma_{q,p}(t)  \;= \;
 \sqrt{-\frac{\hbar}{2}\frac{a_{t}^{\makebox{\tiny I}}}{a_{t}^{\makebox{\tiny R}}}}
\end{equation}
and $\overline{\sigma}_{q,p}$ corresponds to the value of
$\sigma_{q,p}(t)$ when the state $|\psi_{t}\rangle$ is a stationary
Gaussian solution.

After a rather long calculation (see Appendix A for the details), it
is possible to show that:
\begin{eqnarray} \label{eq:loc}
\lefteqn{\frac{d}{dt}\, {\mathbb E} [ \sigma_{O}^2(t) ] \; = \;} &&
\nonumber \\
& = & - 4 \lambda\, {\mathbb E} \left[ \overline{\sigma}_{q}^2\,
\sigma_{O}^2(t) + \overline{\sigma}_{q,p}^4\,
\left(\frac{\sigma_{q}^2(t)}{\overline{\sigma}_{q}^2} -
\frac{\sigma_{q,p}^2(t)}{\overline{\sigma}_{q,p}^2} \right)^2 +
\right. \nonumber \\
& & \qquad + \left. \frac{\hbar^2}{4 \overline{\sigma}_{q}^4}\,
\left( \sigma_{q}^2(t) - \overline{\sigma}_{q}^2 \right)^2 \right]
\leq 0.
\end{eqnarray}
Since $\sigma_{O}^2(t)$ is by definition a positive quantity, the
above equation is consistent if and only if the r.h.s vanishes for
any $\omega \in \Omega$, with the possible exception of a subset
of measure 0. This is particular implies both that
$\sigma_{O}^2(t) \rightarrow 0$ a.s. and that $\sigma_{q}(t)
\rightarrow \overline{\sigma}_{q}$ a.s., which is the desired
result, i.e., the wavefunction converges to a localized solution.

\subsection{The localization probability}

Once proved that Eq.~\eqref{nle} with the choice~\eqref{efa} for the
operator $A$ induces the localization of the wavefunction in space,
it becomes natural to analyze the probability for a localization to
occur within a given interval of the real axis. Such an analysis can
be developed along the same line of Ref.~\cite{bassi-drm}.

Let us consider the probability measure:
\begin{equation} \label{pmes}
\mu_{t}(\Delta) \; \equiv \; {\mathbb E}_{{\mathbb P}}\left[
\|P_{\Delta}\psi_{t}\|^2\right],
\end{equation}
defined on the Borel sigma--algebra ${\mathcal B}({\mathbb R})$ of
the real axis, where $P_{\Delta}$ is the projection operator
associated to the Borel subset $\Delta$ of ${\mathbb R}$; such a
measure is identified by the density $p_{t}(x) \equiv {\mathbb
E}_{{\mathbb P}} [| \psi_{t}(x)|^2 ]$:
\begin{equation}
\mu_{t}(\Delta) \; = \; \int_{\Delta} p_{t}(x) \, dx.
\end{equation}
The density $p_{t}(x)$ corresponds to the diagonal element $\langle
x | \rho_{t} | x \rangle$ of the statistical operator $\rho_{t}
\equiv {\mathbb E}_{{\mathbb P}}\left[
|\psi_{t}\rangle\langle\psi_{t}|\right]$, solution of the
master-equation~\eqref{eq:1}. In Appendix~\ref{sec:gener-solut-mast}
we show how the general solution of this master-equation in the
position representation can be obtained; the final expression, as a
function of the solution of the free Schr\"odinger equation
$\rho_t^{S}$ (i.e., with $\lambda = \alpha = 0)$, is:
\begin{widetext}
\begin{multline}
   \label{eq:33}
   \langle q_1 | \rho_t | q_2 \rangle = \frac{1}{2\pi\hbar}\int dk\int
   dy\,e^{- (i/\hbar)ky} F[k, q_{1} - q_{2}, t]\\ \quad \times \left\langle {
        y+\frac{q_1+q_2}{2}+\frac{q_1-q_2}{2}e^{-2\lambda\alpha
          t}+\frac{kt}{2m} {\left( 1-\frac{\Gamma_{t}}{2\lambda\alpha t}
          \right)} \left|\vphantom{{\left(
                  1-\frac{\Gamma_{t}}{2m\lambda\alpha} \right)}}\rho_t^{S}
           \vphantom{{\left( 1-\frac{\Gamma_{t}}{2m\lambda\alpha}
               \right)}}\right|
        y+\frac{q_1+q_2}{2}-\frac{q_1-q_2}{2}e^{-2\lambda\alpha
          t}-\frac{kt}{2m} {\left( 1-\frac{\Gamma_{t}}{2\lambda\alpha t}
          \right)} } \right\rangle,
\end{multline}
with
\begin{equation}
   F [k,x,t] = \exp
   \left\{{-\frac{\lambda\alpha^2}{2\hbar^2}k^2t} +
   \frac{1}{8\alpha \Gamma_{t}^2} \left[ \left(
            x\,e^{-2\lambda\alpha
              t}-\frac{\Gamma_{t}}{2m\lambda\alpha}k \right)^2K_1 (t)+2
         x\, \left( x\,e^{-2\lambda\alpha
              t}-\frac{\Gamma_{t}}{2m\lambda\alpha}k \right)K_2 (t)+
         x^2K_3 (t) \right] \right\}
\end{equation}
and
\begin{equation}
   \label{eq:34}
   \begin{split}
      K_1 (t)&= \Gamma_{t}^2+2\Gamma_{t}-4\lambda\alpha t, \\
      K_2 (t)&= e^{-4\lambda\alpha t}+4\lambda\alpha t e^{-2\lambda\alpha t}-1, \\
      K_3 (t)&= -4\lambda\alpha t e^{-4\lambda\alpha
        t}-\Gamma_{t}^2+2\Gamma_{t} e^{-2\lambda\alpha t},
\end{split}
\end{equation}
where we have defined $\Gamma_{t}=1-e^{-2\lambda\alpha t}$. Taking
the diagonal matrix elements one has:
\begin{equation}
   \label{eq:35}
   p_t (x) = \frac{1}{2\pi\hbar}\int dk\int
   dy\,e^{- (i/\hbar)ky}\, F[k,0,t]\, \left\langle { y+x+\frac{kt}{2m} {\left(
             1-\frac{\Gamma_{t}}{2\lambda\alpha t} \right)}
        \left|\vphantom{{\left(
                  1-\frac{\Gamma_{t}}{2m\lambda\alpha} \right)}}\rho_t^{S}
           \vphantom{{\left( 1-\frac{\Gamma_{t}}{2m\lambda\alpha}
               \right)}}\right|
        y+x-\frac{kt}{2m} {\left( 1-\frac{\Gamma_{t}}{2\lambda\alpha t}
          \right)} } \right\rangle,
\end{equation}
and
\begin{equation} \label{eq:35bis}
F[k,0,t] = \exp
\left\{-\frac{\lambda\alpha^2}{2\hbar^2}k^2t +
\frac{1}{32m^2\lambda^2\alpha^3}k^2 K_1 (t) \right\}
\end{equation}

According to the values~\eqref{eql} and \eqref{eqa} for $\lambda_0$
and $\alpha_0$, the above expressions can be expanded for
small\footnote{This means that we are considering only times $t \ll
(\lambda \alpha)^{-1} \simeq 10^{20}$ s.} $\lambda\alpha t$, leading
to:
\begin{equation}
   \label{eq:31}
   p_t (x) \, \simeq \, \frac{1}{2\pi\hbar}\int dk\int
   dy\,e^{- (i/\hbar) ky}
\exp \left\{- \left[\frac{\lambda k^2}{6 m^2}t^3 +
\frac{\lambda \alpha^2 k^2}{2 \hbar^2} t \right]\right\}
\left\langle { y+x+ \frac{\lambda\alpha k t^2}{2m}
        \left|\vphantom{{\left(
                  1-\frac{\Gamma_{t}}{2m\lambda\alpha} \right)}}\rho_t^{S}
           \vphantom{{\left( 1-\frac{\Gamma_{t}}{2m\lambda\alpha}
               \right)}}\right|
        y+x-\frac{\lambda\alpha k t^2}{2m}} \right\rangle.
\end{equation}
We now focus on the case of a macroscopic object (let us say
$m\geq 1\ \mathrm{g}$); one can further approximate the above
expression by noting that for such values of $m$ the exponential
factor appearing in Eq.~\eqref{eq:31} damps all matrix elements
such that the term $\lambda \alpha k t^2 /2m$ is not vanishingly
small; e.g. when $\lambda \alpha k t^2 /2m \geq 10^{-15} $ m, then
the second exponential in the above equation is much smaller than
$e^{-10^{15}(m/\makebox{\tiny Kg})}$. We can then neglect the two
terms in the matrix elements and perform the integration over $k$,
and we get:
\begin{equation}
   \label{eq:36}
   p_t (x) \; \simeq \; \sqrt{\frac{\beta_t}{\pi}}\int dy\, e^{-\beta_t y^2} p_t^{S} (x+y)
\end{equation}
with
\begin{equation}
   \label{eq:37}
   \beta_t \; = \; \frac{3}{2\hbar^2}\frac{m^2}{\lambda\left[1+3
   \left(\frac{m\alpha}{\hbar t}\right)^2\right]}\frac{1}{t^3}
   \quad \left\{
   \begin{array}{ll}
   \simeq
   10^{43}\displaystyle \left(\frac{m}{\mathrm{kg}}\right)\left(\frac{\mathrm{sec}}{t}\right)^3
   & \makebox{for: $t \geq 10^{-11}$ s}, \\
   \geq
   10^{65}\displaystyle \left(\frac{m}{\mathrm{kg}}\right)\left(\frac{\mathrm{sec}}{t}\right)
   & \makebox{for: $t \leq 10^{-11}$ s}
   \end{array}
   \right.
\end{equation}
\end{widetext}
The exponent in~\eqref{eq:36} is extremely peaked with respect to
the typical values the probability density $p_t^{S} (x)$ associated
to the wavefunction of a macroscopic object takes, so that with very
high accuracy we have:
\begin{equation}
   \label{eq:38}
   p_t (x) \; \simeq \; p_t^{S} (x).
\end{equation}
As discussed in ref.~\cite{bassi-drm}, the probability measure
$\mu_{t}(\Delta)$ which we have shown to be extremely close to the
quantum probability obtained from the free Schr\"odinger equation
can be interpreted as a probability measure for the collapse of
the wavefunction of the macroscopic object within $\Delta$, when
$\Delta$ corresponds to an interval of the real axis of width
greater or equal e.g. to $10^{-5}$ cm~\cite{bassi-drm}.

To conclude, the previous analysis shows that under the above listed
conditions the probability for the wavefunction of a macro-object to
be localized within an interval of the real axis is almost equal to
the corresponding quantum probability as given by the Born rule.

\section{Dynamics of microscopic and macroscopic systems}
\label{sec:micro-macro}

In this section we discuss how our reduction model is related both
to quantum and to classical mechanics. Our aim is to show that at
the microscopic level the physical predictions of the model are
almost identical to standard quantum predictions and that, at the
same time, the model with high accuracy reproduces classical
mechanics at the macro--level.

\subsection{Micro--systems: comparison with standard quantum
mechanics} \label{sec:micro}

Microscopic system cannot be directly observed, and their properties
can be analyzed only by resorting to suitable measurement
procedures. All physical predictions of our model, concerning the
outcome of measurements, have the form ${\mathbb E}_{\mathbb P} [
\langle \psi_{t}| S | \psi_{t} \rangle ]$ where $S$ is a suitable
self--adjoint operator, typically a projection operator and it is
easy to show that ${\mathbb E}_{\mathbb P} [ \langle \psi_{t}| S |
\psi_{t} \rangle ] \equiv \makebox{Tr}[S \rho_{t}]$, where the
statistical operator $\rho_{t} \equiv {\mathbb E}_{\mathbb P} [ |
\psi_{t}\rangle \langle \psi_{t}| ]$ satisfies Eq. (\ref{eq:1}).
Accordingly, as already discussed in Sec.~\ref{sec:mast-equat-stat},
the testable effects of the stochastic process on the wavefunction
are similar to the effect induced by a quantum environment on the
particle, when both friction and diffusion are taken into account.

With our choice (\ref{eql}) and (\ref{eqa}) for $\lambda_{0}$ and
$\alpha_{0}$, the testable effects of the stochastic process are
of the same order of magnitude of those induced by the interaction
of the system with particles and radiation of the {\it
intergalactic space} \cite{dec}: such effects are very small and
masked by most other sources of decoherence, so that they can be
tested only by resorting to sophisticated experiments
\cite{exp,ex1}. This implies that the physical predictions of our
model are very close to standard quantum mechanical predictions.

\subsection{Macro--objects: comparison with classical mechanics}
\label{sec:macro}

A macroscopic object is made of elementary constituents strongly
interacting among each other and, according to our model, its
dynamics is governed by the following stochastic differential
equation, which is the straightforward generalization of Eq.
(\ref{nle}) to a system of $N$ particles:
\begin{eqnarray} \label{nlenp}
\lefteqn{ d\,\psi_{t}(\{x\})  = \left[ -\frac{i}{\hbar}\,
H_{\makebox{\tiny TOT}}\, dt + \sum_{n=1}^{N}
\sqrt{\lambda_{n}}\, \left( A_{n} - r_{n t}\right) dW^{n}_{t}\right.}
& & \nonumber \\
& & \qquad - \left. \sum_{n=1}^{N}\frac{\lambda_{n}}{2} \left(
A_{n}^\dag A_{n}^{\phantom \dag} - 2 r_{nt} A_{n} + r_{nt}^2
\right) dt\right]
\psi_{t}(\{x\}); \nonumber \\
& &
\end{eqnarray}
the symbol $\{ x \} \equiv x_{1}, x_{2}, \ldots , x_{N}$ represents
the $N$ spatial coordinates of the configuration space of the
composite system; $W^{1}_{t}, W^{2}_{t}, \ldots W^{N}_{t}$ are $N$
independent Wiener processes; $r_{n t} = \langle \psi_{t}|
[A^{\phantom \dagger}_{n} + A^{\dagger}_{n}] |\psi_{t} \rangle/2$
and the localization operators $A_{n}$ are given by expressions
(\ref{efa}), with $q$ replaced by $q_{n}$, the position operator of
the $n$--th particle, and $p$ replaced by $p_{n}$, the corresponding
operator. Furthermore:
\begin{equation} \label{parn}
\lambda_{n} = \frac{m_{n}}{m_{0}}\, \lambda_{0} \qquad \alpha_{n} =
\frac{m_{0}}{m_{n}}\, \alpha_{0}
\end{equation}
and
\begin{equation}
H_{\makebox{\tiny TOT}} \; = \; H_{\makebox{\tiny TOT}}^{0} \, +
\, \sum_{n=1}^{N} \frac{\lambda_{n} \alpha_{n}}{2} \, \left\{
q_{n}, p_{n} \right\},
\end{equation}
$H_{0}$ being the standard quantum Hamiltonian for the composite
system.

As custom, we separate the motion of the center of mass from the
internal motion. To this end let:
\begin{equation}
Q \; = \; \displaystyle \frac{1}{M}\, \sum_{n=1}^{N} m_{n}\, q_{n}
\qquad \left(M \; = \; \sum_{n=1}^{N} m_{n}\right)
\end{equation}
be the position operator associated to the center-of-mass coordinate
$R$, and $\tilde{q}_{n}$ the position operators associated to the
internal coordinates $\tilde{x}_{n} = x_{n} - R$ ($n = 1, \ldots ,
N$); let also $P$ and $\tilde{p}_{n}$ be the corresponding momentum
operators. Then, if $H_{\makebox{\tiny TOT}}^{0}$ can be written as
the sum of a term $H_{\makebox{\tiny CM}}^{0}$ associated to the
center of mass and a term $H_{\makebox{\tiny REL}}^{0}$ associated
to the internal motion, it is easy to prove that the dynamics of the
two types of degrees of freedom decouple; in particular the equation
for the center of mass
--- the only one we consider here --- becomes:
\begin{eqnarray} \label{nlecm}
\lefteqn{ d\,\psi_{t}(R) = \left[ -\frac{i}{\hbar}\,
H_{\makebox{\tiny CM}}\, dt + \sqrt{\lambda_{\makebox{\tiny
CM}}}\, \left( A_{\makebox{\tiny CM}} - r_{\makebox{\tiny
CM,t}}\right) dW_{t}\right.} & & \nonumber \\
& & - \left. \frac{\lambda_{\makebox{\tiny CM}}}{2}\, \Big(
A_{\makebox{\tiny CM}}^\dag A_{\makebox{\tiny CM}} - 2
r_{\makebox{\tiny CM,t}} A_{\makebox{\tiny CM}} +
r_{\makebox{\tiny CM,t}}^2 \Big) dt\right] \psi_{t}(R), \qquad
\end{eqnarray}
with:
\begin{eqnarray}
H_{\makebox{\tiny CM}} & = & H_{\makebox{\tiny CM}}^{0} +
\frac{\lambda_{\makebox{\tiny CM}}\alpha_{\makebox{\tiny CM}}}{2}
\left\{ Q, P \right\},  \\
r_{\makebox{\tiny CM,t}} & = & \frac{1}{2}\, \langle \psi_{t}| [
A^{\dagger}_{\makebox{\tiny CM}} + A^{\phantom
\dagger}_{\makebox{\tiny CM}}] |\psi_{t} \rangle, \quad
\\
A_{\makebox{\tiny CM}} & = & Q \, + \, i
\frac{\alpha_{\makebox{\tiny CM}}}{\hbar}\, P,
\end{eqnarray}
and $\lambda_{\makebox{\tiny CM}}$ and $\alpha_{\makebox{\tiny CM}}$
defined by Eqs. (\ref{eqp}), with $m$ equal to the total mass $M$ of
the composite system. Note that the separation of the center-of-mass
motion from the relative motion, for the non-Schr\"odinger terms of
Eq.~\eqref{nlenp}, is possible because of the specific dependence of
the parameters $\lambda_{n}$ and $\alpha_{n}$ on the masses $m_{n}$
of the particles as given by Eq.~\eqref{parn}.

According to Eq. (\ref{nlecm}) the center of mass behaves like a
particle whose dynamics, in the free case, has been discussed in
detail in Secs. \ref{sec:1gauss} and \ref{sec:asym-beh}; we now show
how the large numerical value for $M$, typical of macroscopic
objects, affects the time evolution of the center-of-mass
wavefunction.
\\ \\
\noindent {\it 1 Collapse rate.} According to Eq.~\eqref{eq:loc},
and assuming that the wavefunction is not already localized in
space, i.e., that $\sigma_{q}(t) \gg \overline{\sigma}_{q}$, one has:
\begin{eqnarray} \label{eq:macro-red-1}
\left| \frac{d}{dt}\, {\mathbb E} [ \sigma_{O}^2(t) ] \right| & \geq
& \frac{\lambda \hbar^2}{\overline{\sigma}_{q}^4}\, \sigma_{q}^4(t)
\; =
\nonumber \\
& = & \left( \frac{2 \lambda_{0} \omega^2 \sin^2 \theta}{m_{0}}
\right)\, m^3\, \sigma_{q}^4(t), \label{eq:macro-red-2}
\end{eqnarray}
with $2 \lambda_{0}  \omega^2 \sin^2 \theta / m_{0}  \simeq 10^{15}$
Kg$^{-1}$ m$^{-2}$ sec$^{-3}$. In the microscopic case the r.h.s. of
Eq.~\eqref{eq:macro-red-2} is in general small and negligible as we
expect it to be, since the reduction mechanism must be ineffective
on microscopic systems; in the macroscopic case instead it rapidly
becomes big, due to the large value of the mass $m$, ensuring that
{\it any} initial wavefunction rapidly converges towards a localized
solution. We can then assume that, possibly after a very short
transient period, the wavefunction describing the motion of the
center of mass of any macro--object is practically localized in
space.
\\ \\
\noindent {\it 2 Behavior of the stationary solution for macroscopic
objects.} We now discuss the time evolution of a typical localized
wavefunction, i.e., a Gaussian stationary solution. Eqs.~\eqref{ep}
and~\eqref{em} imply that the two maxima in position and momentum of
such wavefunctions fluctuate around their mean values; we now show
that in the macroscopic regime these fluctuations are extremely
small.

As a matter of fact Eqs.~\eqref{eq1c} to~\eqref{eq3c} imply for a
stationary solution:
\begin{eqnarray}
C_{q^2}(t) &=& \lambda\ell^2 t -
\frac{\hbar\ell}{2\lambda\alpha^2m}\left( \frac{\cos\theta -
\kappa}{\sin\theta} \right)\left( 1 - e^{-2\lambda\alpha t}\right)
\nonumber \\&+& \frac{\hbar^2}{16\lambda^2\alpha^3m^2}\left(
\frac{\cos\theta -
\kappa}{\sin\theta} \right)^2\left( 1 - e^{-4\lambda\alpha t}\right) \\
C_{p^2}(t) &=& \frac{\hbar^2}{4\alpha}\left( \frac{\cos\theta -
\kappa}{\sin\theta} \right)^2\left( 1 - e^{-4\lambda\alpha
t}\right),
\end{eqnarray}
where $\ell$ is defined as follows:
\begin{equation}
\ell \; = \; -\frac{\hbar}{2\lambda\alpha m}\left(
\frac{\cos\theta - \kappa}{\sin\theta} \right) - \left(
\frac{\sqrt 2 \hbar}{m\omega\sin\theta} \!\!- \alpha\right).
\end{equation}
Since the exponential factors decay very slowly (we remember that
$\lambda \alpha \simeq 10^{-20}$ sec${}^{-1}$) it is physically
significant to consider only the linear term of their Taylor
expansion; one then gets:
\begin{eqnarray}
C_{q^2}(t) & \simeq & \frac{4\hbar^2\lambda_0}{m_0
\omega^2}\frac{t}{m} \; \simeq \; 10^{-33} \left( \frac{\rm Kg}{m}
\right)
\left(\frac{t}{\rm sec}\right) {\rm m^2} \\
C_{p^2}(t) & \simeq & \frac{\hbar^2\lambda_0}{m_0} m t \; \simeq
\; 10^{-43} \left( \frac{m}{\rm Kg} \right)\left(\frac{t}{\rm
sec}\right)
\frac{\rm Kg^2 m^2}{\rm sec^2} \nonumber \\
& &
\end{eqnarray}
which are very small quantities, when $m$ is the mass of a
macro-object. Accordingly, in the macroscopic case the actual value
of the two peaks in position and momentum of a stationary Gaussian
solution are very close to their average values which, as we have
seen, evolve in time according to Newton's laws for a free particle
moving in a (very weakly) dissipative medium. This proves that with
high accuracy a stationary solutions for the center of mass of a
macro--system practically behave like a point moving in space
according to the laws of classical mechanics.

\section{Time evolution of the mean energy}
\label{sec:nrg}

We now discuss one of the main purposes of our work, i.e., we show
that within our collapse model the energy of an isolated system does
not increase with a constant (even if negligible) rate, but reaches
an asymptotic finite value. Indeed, this result is entailed by Eq.
(\ref{eq:1}) for the statistical operator, which motivated our
choice for the localization operator $A$.

As a matter of fact, the mean value ${\mathbb E}[\langle H_0
\rangle_{t}]$ of the energy satisfies the following equation:
\begin{equation}
\frac{d}{dt}\, {\mathbb E}[\langle H_0 \rangle_{t}] \; = \;
\frac{\lambda \hbar^2}{2m} \, - \, 4\lambda\alpha{\mathbb E}[\langle
H_0 \rangle_{t}].
\end{equation}
whose solution is:
\begin{equation}
{\mathbb E}[\langle H_0 \rangle_{t}] = \left(E_0 -
\frac{\hbar^2}{8m\alpha}\right)e^{-4\lambda\alpha t} +
\frac{\hbar^2}{8m\alpha}.
\end{equation}
As we see, the mean energy of an isolated system changes in time and
thus is not conserved; anyway it does not diverge for $t \rightarrow
+ \infty$, but reaches the asymptotic finite value:
\begin{equation}
\lim_{t\to +\infty}{\mathbb E}[\langle H_0 \rangle_{t}] =
\frac{\hbar^2}{8m\alpha}\; = \; \frac{\hbar^2}{8m_{0}\alpha_{0}},
\end{equation}
corresponding to a temperature
\begin{equation} \label{eq:temp}
T \; = \; \frac{\hbar^2}{4 m_{0}\alpha_{0} k_{B}} \; \simeq \;
10^{-1}\; \makebox{K},
\end{equation}
which is independent of the mass of the particle. Note that the time
evolution of ${\mathbb E}[\langle H_0 \rangle_{t}]$ is very slow,
the rate of change being equal to $4\lambda\alpha =
4\lambda_{0}\alpha_{0} \simeq 10^{-20} \; \makebox{sec$^{-1}$}$
which means that, with very high accuracy, the mean energy remains
constant for very long times.

The above equations imply that the stochastic process acts like a
dissipative medium which, due to friction, slowly thermalizes all
systems to the temperature $T$ by absorbing or transferring energy
to them depending on their initial state. Note that according to
Eq.~\eqref{eq:temp} a very ``cold'' medium is enough to guarantee
the localization in space of the wavefunctions of macroscopic
objects. Note also that one recovers a GRW-type equation by
setting $\alpha \rightarrow 0$, which corresponds to the high
temperature limit $T \rightarrow + \infty$. This implies that the
reason why in the original GRW reduction model \cite{drm1} the
energy increases and eventually diverges is that the noise acts
like an infinite-temperature stochastic medium. Our model also
shows that the increase of the mean energy of a quantum system
subject to spontaneous localizations is not an intrinsic feature
of space-collapse models (indeed, according to our model, the mean
energy {\it decreases} for typical quantum systems) and it can be
(partly) avoided with a suitable choice of the localizations
operators.

The above discussion suggests that the model can be further
developed by promoting $W_{t}$ to a {\it real physical} medium with
its own equations of motions, having a stochastic behavior to which
a temperature $T$ can be associated and such that, with good
accuracy, can be treated like a Wiener process. This would imply not
only that the medium acts on the wavefunction by changing its state
according to Eq.~\eqref{nle}, but also that the wavefunction acts
back on the medium according to equations which still have to be
studied. The above suggestion opens the way to the possibility that
by taking into account the energy of both the quantum system {\it
and} the stochastic medium one can restore perfect energy
conservation not only on the average but also for single
realizations of the stochastic process. A similar proposal has been
considered by P. Pearle~\cite{pen} and by S. Adler~\cite{ad3}.

These ideas will be subject of future research; we conclude by
noting that, whatever its nature can be, the stochastic medium
cannot be quantum in the usual sense since its coupling to the
particle is not a standard coupling between two quantum systems:
Eq.~\eqref{nle}, in fact, is {\it not} the standard Schr\"odinger
equation with a stochastic potential.

\section{Conclusions}
\label{concl} We have presented and analyzed a collapse--model
which preserves the standard quantum mechanical predictions and
reproduces classical mechanics at the macroscopic level, at the
same time avoiding the infinite growth of energy of the system, a
criticized feature \cite{Ballentine-Gallis} of the space collapse
models appeared in the literature. 
This has been obtained by
drawing on an analogy with quantum Brownian motion, where friction
effects, directly related to diffusion for the preservation of
Heisenberg  uncertainty relation, guarantee that the energy of the test particle reaches a
finite value depending on the parameters of the model. 
The model
is also characterized by the fact that the related master-equation
complies with the general structure of translation-covariant
generators of quantum-dynamical semigroups obtained by Holevo
\cite{HolevoJMP-HolevoRMP32-HolevoRMP33-HolevoRAN}, so that
symmetry under translation is correctly kept into account, as
compulsory for dynamical reduction models, which should not
introduce any preferred space location.
\par
Needless to say the exploited analogy with the quantum Brownian motion
master-equation, typically used for the description of dissipation and
decoherence, should by no way induce confusion on the different nature
of the two issues of decoherence and of the measurement or
macro--objectification problem in quantum mechanics; as it has been
stressed also in recent publications
\cite{dec,JoosLNPH99-AnastopoulosIJTP02- AdlerSHPMP03} decoherence
does not provide a solution to the measurement problem. This important
conceptual difference notwithstanding, dynamical reduction models and
models of environmental decoherence both impinge on the same
mathematical inventory, typically used in the theory of open quantum
system \cite{Petruccione}, so that results obtained in the one field
can often be fruitfully exploited in the other.
\par
An extension of this approach might be pursued in order to cope with
the infinite energy growth also in the original GRW model of
dynamical reductions \cite{drm1}, building on the analogy with the
quantum linear Boltzmann equation \cite{art7}.

\section*{Acknowledgements}

This work was partially supported by INFN and by MIUR under
Cofinanziamento and FIRB. The work of A.B. was supported by the EU
grant No. 507 547 93 - 1, AOST. Nr. 851601 - 5. We thank S.~L.~Adler
for useful suggestions on a draft of the manuscript.

\appendix
\section{The mean rate of change of the variance $\sigma_O$}\label{ZoupasCalculation}

In order to derive Eq. \eqref{eq:loc}, we first compute the average
value of the time derivative of the second moments of the operators
$q$ and $p$ and of their symmetrized correlation. Using the
evolution equation \eqref{nle}, we obtain through It\^o calculus:
\begin{eqnarray}
\frac{d}{d t}\, {\mathbb E} [ \sigma_{q}^2(t) ]\phantom{\:\,\:\!} &
= &
\frac{2 {\mathbb E}[\sigma_{q,p}^2(t)]}{m} - 4\lambda {\mathbb E} [\sigma_{q}^4(t)]
+ 4\alpha\lambda {\mathbb E} [\sigma_{q}^2(t)], \nonumber \\
& &\label{a5-1}\\
\frac{d}{d t}\, {\mathbb E} [ \sigma_{p}^2(t) ] \phantom{\:\,\:\!} &
= & -4\lambda {\mathbb E} [\sigma_{q,p}^4(t)] -4\alpha\lambda
{\mathbb E} [\sigma_{p}^2(t)]
+ \lambda\hbar^2, \nonumber \\
& &\label{a5-2}\\
\frac{d}{d t}\, {\mathbb E} [ \sigma_{q,p}^2(t) ] & = &
\frac{{\mathbb E} [\sigma_{p}^2(t)]}{m} - 4\lambda {\mathbb E}
[\sigma_{q,p}^2(t)\sigma_{q}^2(t)] \label{a5-3}.
\end{eqnarray}
Since $a_{t}$ is constant for a stationary solution, the stationary
values of variances and correlation are such that the right-hand
sides of \eqref{a5-1} to \eqref{a5-3} vanish:
\begin{eqnarray}
&&\frac{\overline{\sigma}_{q,p}^2}{m}-2\lambda\overline{\sigma}_{q}^4+2\alpha\lambda\overline{\sigma}_{q}^2\;=\;0,
\label{a5-4}\\
&&\overline{\sigma}_{q,p}^4+\alpha\overline{\sigma}_{p}^2-\frac{\hbar^2}{4}\;=\;0,\label{a5-5}\\
&&\frac{\overline{\sigma}_{p}^2}{m}-4\lambda\overline{\sigma}_{q,p}^2\overline{\sigma}_{q}^2\;=\;0\label{a5-6}.
\end{eqnarray}
Moreover Eq. \eqref{pin} can be rewritten as: \be
\overline{\sigma}_{q}^2\overline{\sigma}_{p}^2 -
\overline{\sigma}_{q,p}^4 \;=\;\frac{\hbar^2}{4}. \label{a5-7}\ee

For convenience we define the quantity $X,Y {\rm and\ }Z$ by the
equations:
\begin{eqnarray*}
\sigma_{q}^2(t) & = &\overline{\sigma}_{q}^2(1+X),\\
\sigma_{p}^2(t) & = &\overline{\sigma}_{p}^2(1+Y),\\
\sigma_{q,p}^2(t) & = &\overline{\sigma}_{q,p}^2(1+Z),
\end{eqnarray*}
so that Eq. \eqref{AO2} becomes
\begin{equation}\label{a5-8}
\sigma_{O}^2(t) \; = \;\overline{\sigma}_{p}^2(X+Y) -
\frac{2\overline{\sigma}_{q,p}^4}{\overline{\sigma}_{q}^2}Z,
\end{equation}
where we also made use of the Eq. \eqref{a5-7}.

Using Eqs. \eqref{a5-1} to \eqref{a5-3} we obtain for the average
value of the time derivative of ${\sigma}_{O}^2$:
\begin{multline}
\frac{d}{d t}\, {\mathbb E} [ \sigma_{O}^2(t) ] =
w_1 {\mathbb E} [X] + w_2 {\mathbb E} [Y] + w_3 {\mathbb E} [Z]\\
-4\lambda\left(\overline{\sigma}_{q}^2\overline{\sigma}_{p}^2
{\mathbb E} [X^2] +\overline{\sigma}_{q,p}^4 {\mathbb E} [Z^2]
-2\overline{\sigma}_{q,p}^4 {\mathbb E} [XZ]\right)\label{eq:8}
\end{multline}
with
\begin{eqnarray}
w_1 & = &
-8\lambda\left(\overline{\sigma}_{q}^2\overline{\sigma}_{p}^2 -
\frac{1}{2}\alpha\overline{\sigma}_{p}^2 - \overline{\sigma}_{q,p}^4\right),\label{a5-9}\\
w_2 & = & -2\left(2\alpha\lambda\overline{\sigma}_{p}^2 +
\frac{\overline{\sigma}_{q,p}^2}{m}\frac{\overline{\sigma}_{p}^2}{\overline{\sigma}_{q}^2}\right),\label{a5-10}\\
w_3 & = &
2\frac{\overline{\sigma}_{q,p}^2}{m}\frac{\overline{\sigma}_{p}^2}{\overline{\sigma}_{q}^2}.\label{a5-11}
\end{eqnarray}
From Eqs. \eqref{a5-7} and \eqref{a5-5} we get for $w_1$ \be w_1 \;
= \; -4\lambda\overline{\sigma}_{q}^2\overline{\sigma}_{p}^2,
\label{a5-12} \ee while using Eq. \eqref{a5-4} one can prove that
\be w_2 \; =\; w_1. \ee Finally, from Eq. \eqref{a5-6}, we find that
$w_3$ and $w_1$ are related as: \be w_3 \; =\;
-\frac{2\overline{\sigma}_{q,p}^4}{\overline{\sigma}_{q}^2\overline{\sigma}_{p}^2}w_1
\ee so that we have \be w_1X+w_2Y+w_3Z \; = \; w_1\left(
X+Y-\frac{2\overline{\sigma}_{q,p}^4}{\overline{\sigma}_{q}^2\overline{\sigma}_{p}^2}Z\right)\;
= \; -4\lambda\overline{\sigma}_{q}^2\sigma_{O}^2(t), \ee where the
last equality is obtained through Eqs. \eqref{a5-8} and
\eqref{a5-12}. Finally, the above result together with Eq.
\eqref{a5-7} allow us to write
\begin{equation}
\frac{d}{d t}\, {\mathbb E} [ \sigma_{O}^2(t) ] \; = \; -4\lambda
{\mathbb E}\left[\overline{\sigma}_{q}^2\sigma_{O}^2(t)+
\overline{\sigma}_{q,p}^4(X-Z)^2 + \frac{\hbar^2}{4}X^2 \right],
\end{equation}
which is just Eq. \eqref{eq:loc}.

\section{General solution of the master-equation}
\label{sec:gener-solut-mast}

We now show how to obtain the general solution of the
master-equation
\begin{eqnarray} \label{eq:1tris}
\frac{d}{dt}\, \rho_t &= & - \frac{i}{\hbar}\, \left[ H_0, \rho_t
\right] \, - \frac{\lambda}{2}\, \left[q, \left[ q, \rho_t \right]
\right] \nonumber \\
& & - \frac{\lambda \alpha^2}{2\hbar^2}\, \left[p, \left[ p,
\rho_t \right] \right]\, - \, i\, \frac{\lambda \alpha}{\hbar}\,
\left[ q, \left\{ p, \rho_t \right\} \right],
\end{eqnarray}
given in~\eqref{eq:1}, which has the same form of the quantum
Brownian motion master-equation, partially following an appendix
of \cite{dec}. In particular we want to express the general
solution in the position representation $\langle q_1 | \rho_t |
q_2 \rangle$ as a function of the solution of the pure
Schr\"odinger equation $\langle q_1 | \rho_t^{S} | q_2 \rangle$,
in order to point out the corrections to the position probability
density due to the non-linear stochastic modification of the
Schr\"odinger dynamics.

As a first step we want to express the solution
of~\eqref{eq:1tris} as a function of the generic initial condition
$\rho_0$ according to
\begin{equation}
   \label{eq:9}
   \langle q_1 | \rho_t | q_2 \rangle = \int dq_{10} dq_{20}\,
  G (q_1,q_2,t;q_{10},q_{20},0)
\langle q_{10} |\rho_0 |q_{20}\rangle,
\end{equation}
where $G$ is the Green function solution of the partial
differential equation associated to~\eqref{eq:1tris} in the
position representation satisfying the following initial condition
\begin{equation}
   \label{eq:10}
   G (q_1,q_2,t;q_{10},q_{20},0) \xrightarrow[{t\rightarrow 0}]{}
   \delta (q_1 -q_{10})\delta (q_2-q_{20}).
\end{equation}
Once $G$ is known one may immediately express the general solution
as a function of the solution of the Schr\"odinger equation by
means of the free propagator $G_0$
\begin{multline}
   \label{eq:11}
   \langle q_1 | \rho_t | q_2 \rangle = \int dq_{10} dq_{20}\int
   du dv\,
  G (q_1,q_2,t;q_{10},q_{20},0) \\ \times G_0(q_{10},q_{20},0;u,v,t)
\langle u |\rho_t^{S} |v\rangle .
\end{multline}
For calculational purposes it is however convenient to consider
the quantity
\begin{equation}
   \label{eq:12}
   \tilde{\rho}_t (k,x)=\Tr \left(\rho_t e^{\frac{i}{\hbar} (kq+xp)}\right),
\end{equation}
corresponding to the characteristic function associated to the
Wigner function. The master-equation \eqref{eq:1tris} thus becomes
\begin{multline}
   \label{eq:13}
  \frac{\partial}{\partial t}\tilde{\rho}_t
  (k,x)=\frac{k}{m}\frac{\partial}{\partial x}\tilde{\rho}_t
  (k,x)-\frac{\lambda}{2}x^2\tilde{\rho}_t
  (k,x)\\-\frac{\lambda\alpha^2}{2\hbar^2}k^2\tilde{\rho}_t (k,x)-
  2\lambda\alpha x\frac{\partial}{\partial x}\tilde{\rho}_t (k,x),
\end{multline}
while~\eqref{eq:9} and~\eqref{eq:11} take the form
\begin{equation}
   \label{eq:14}
   \tilde{\rho}_t (k,x)=\int dk_0 dx_0\, \tilde{G} (k,x,t;k_0,x_0,0)\tilde{\rho}_0 (k_0,x_0)
\end{equation}
and
\begin{multline}
   \label{eq:15}
   \tilde{\rho}_t (k,x)=\int dk_0 dx_0\int dr ds\, \tilde{G}
   (k,x,t;k_0,x_0,0)\\ \times\tilde{G}_0 (k_0,x_0,0; r,s,t)\tilde{\rho}_t^{S} (r,s)
\end{multline}
respectively, where $\tilde{\rho}_t^{S} (r,s)$ is again the
solution of the free Schr\"odinger equation, $\tilde{G}_0$ is
simply given by
\begin{equation}
   \label{eq:16}
   \tilde{G}_0 (k,x,t;k_0,x_0,0)=\delta (k-k_0)\delta \left(x-x_0+\frac{k_0}{m}t\right),
\end{equation}
and $\tilde{G}$ satisfies the following initial condition:
\begin{equation}
   \label{eq:17}
   \tilde{G} (k,x,t;k_0,x_0,0)\xrightarrow[{t\rightarrow 0}]{}\delta (k-k_0)\delta (x-x_0).
\end{equation}
Eq.\eqref{eq:15} can be brought back to~\eqref{eq:11} by exploiting
the inverse relation of Eq.~\eqref{eq:12}, i.e.,
\begin{equation}
   \label{eq:18}
   \langle q_1 | \rho_t | q_2 \rangle = \frac{1}{2\pi\hbar}\int dk \,
   e^{-\frac{i}{\hbar}k\left(\frac{q_1+q_2}{2}\right)} \tilde{\rho}_t (k,q_1-q_2).
\end{equation}
The key observation in order to determine $\tilde{G}$ is the fact
that Eq.\eqref{eq:1tris} preserves Gaussian states, so that given
the Gaussian Ansatz:
\begin{equation}
   \label{eq:19}
   \tilde{\rho}_t (k,x)=\exp \left\{{-c_1 k^2 -c_2 kx -c_3 x^2 -ic_4 k -i c_5 x
     - c_6}\right\}
\end{equation}
one easily obtains the evolved state by expressing the
coefficients at time $t$, $c_i (t)$, as a function of the initial
state characterized by the values of the coefficient at time zero.
Considering an initial Gaussian state
$\tilde{\rho}_0^{k_0x_0,\epsilon\eta} (k,x)$ which in the limit of
small $\epsilon$ and $\eta$ approximates the Dirac delta,
according to
\begin{equation}
   \label{eq:20}
   \tilde{\rho}_0^{k_0x_0,\epsilon\eta} (k,x) \xrightarrow[{\epsilon,\eta\rightarrow 0}]{}\delta (k-k_0)\delta (x-x_0)
\end{equation}
one has
\begin{equation}
   \label{eq:21}
   \tilde{G} (k,x,t;k_0,x_0,0)=\lim_{{\epsilon,\eta\rightarrow 0}}\tilde{\rho}_t^{k_0x_0,\epsilon\eta} (k,x).
\end{equation}
Coming back to the Ansatz~\eqref{eq:19}, the coefficients satisfy
the equations
\begin{widetext}
\begin{equation}
   \label{eq:22}
\begin{split}
   \dot{c}_1 (t) &= \frac{c_2 (t)}{m}+ \frac{\lambda\alpha^2}{2\hbar^2} \\
   \dot{c}_2 (t) &=  \frac{2c_3 (t)}{m}-2\lambda\alpha c_2 (t)\\
   \dot{c}_3 (t) &= \frac{\lambda}{2}-4\lambda\alpha  c_3 (t)\\
   \dot{c}_4 (t) &=  \frac{c_5 (t)}{m}\\
   \dot{c}_5 (t) &=  -2\lambda\alpha c_5 (t)
\end{split}
\end{equation}
with solutions
\begin{equation}
   \label{eq:23}
   \begin{split}
      c_1 (t)&= c_1 (0)+c_2 (0)\frac{\Gamma_{t}}{2m\lambda\alpha}+c_3
      (0)\frac{\Gamma_{t}^2}{4m^2\lambda^2\alpha^2}
-\frac{1}{32
        m^2\lambda^2\alpha^3}(\Gamma_{t}^2 + 2\Gamma_{t}-4\lambda\alpha
         t)+\frac{\lambda\alpha^2}{2\hbar^2}t\\
      c_2 (t)&=  c_2 (0)e^{-2\lambda\alpha t}+c_3 (0)\frac{\Gamma_{t} e^{-2\lambda\alpha t}}{m\lambda\alpha}+\frac{\Gamma_{t}^2}{8m\lambda\alpha^2}\\
      c_3 (t)&= \frac{1}{8\alpha}+\left(c_3 (0)-\frac{1}{8\alpha}\right)e^{-4\lambda\alpha t}\\
      c_4 (t)&=  c_4 (0)+c_5 (0)\frac{\Gamma_{t}}{2m\lambda\alpha}\\
      c_5 (t)&= c_5 (0)e^{-2\lambda\alpha t},
\end{split}
\end{equation}
with $\Gamma_{t}=1-e^{-2\lambda\alpha t}$ and $c_6$ simply a
constant. For the choice
\begin{equation}
   \label{eq:24}
   \tilde{\rho}_0^{k_0x_0,\epsilon\eta} (k,x)
   =\frac{1}{\pi\sqrt{\epsilon\eta}} \exp \left\{{-\frac{1}{\epsilon}
        (k-k_0)^2}\right\}
 \exp \left\{{-\frac{1}{\eta} (x-x_0)^2}\right\}
\end{equation}
one has, exploiting~\eqref{eq:23}
\begin{multline}
   \label{eq:25}
   \tilde{\rho}_t^{k_0x_0,\epsilon\eta}
   (k,x)=\frac{1}{\pi\sqrt{\epsilon\eta}}\exp \left\{{-\frac{1}{\epsilon}
     (k-k_0)^2}\right\}
\exp \left\{{-\frac{1}{\eta}\left[\frac{\Gamma_{t}
          k}{2m\lambda\alpha}
        -(x_0-xe^{-2\lambda\alpha t})\right]^2}\right\}\\ \times
   \exp \left\{{-\frac{\lambda\alpha^2}{2\hbar^2}\, k^2 t}\right\}
\exp \left\{{\frac{1}{8\alpha}\left[k^2 \frac{K_1 (t)}{4m^2\lambda^2\alpha^2}
     - kx\frac{\Gamma_{t}^2}{m\lambda\alpha}-x^2 (1-e^{-4\lambda\alpha t})\right]}\right\}
\end{multline}
so that taking the limit one has
\begin{equation}
   \label{eq:27}
   \tilde{G} (k,x,t;k_0,x_0,0)=\delta (k-k_0)\delta \left(\frac{\Gamma_{t}
        k}{2m\lambda\alpha}
      -(x_0-xe^{-2\lambda\alpha t})\right)
   \exp \left\{{-\frac{\lambda\alpha^2}{2\hbar^2}\, k^2 t}\right\}\exp \left\{{\frac{1}{8\alpha\Gamma_{t}^2}[x_0^2
     K_1 (t)+2xx_0K_2 (t)+x^2K_3 (t)]}\right\}
\end{equation}
with
\begin{equation}
   \label{eq:26}
   \begin{split}
      K_1 (t)&= \Gamma_{t}^2+2\Gamma_{t}-4\lambda\alpha t \\
      K_2 (t)&= e^{-4\lambda\alpha t}+4\lambda\alpha t e^{-2\lambda\alpha t}-1\\
      K_3 (t)&= -4\lambda\alpha t e^{-4\lambda\alpha
        t}-\Gamma_{t}^2+2\Gamma_{t} e^{-2\lambda\alpha t}
\end{split}
\end{equation}
and therefore~\eqref{eq:15} now explicitly becomes
\begin{multline}
   \label{eq:28}
   \tilde{\rho}_t
   (k,x)=\exp \left\{{-\frac{\lambda\alpha^2}{2\hbar^2}\, k^2 t}\right\}
   \exp \left\{{\frac{1}{8\alpha\Gamma_{t}^2}\left[\left(xe^{-2\lambda\alpha
             t}+\frac{\Gamma_{t} k}{2m\lambda\alpha}\right)^2 K_1
        (t)+2x\left(xe^{-2\lambda\alpha t}+\frac{\Gamma_{t}
             k}{2m\lambda\alpha}\right)K_2 (t)+x^2K_3
        (t)\right]}\right\}\\ \times
   \tilde{\rho}_t^{S}\left(k, xe^{-2\lambda\alpha t}+\frac{\Gamma_{t}
        k}{2m\lambda\alpha}\left(1-\frac{2\lambda\alpha
           t}{\Gamma_{t}}\right)\right).
\end{multline}
Exploiting the inversion formula~\eqref{eq:18} together with the
expression
\begin{equation}
   \label{eq:29}
   \tilde{\rho}_t
   (k,x)=\int dy\, e^{\frac{i}{\hbar}ky}\langle y+
\frac{x}{2}|\rho_t | y- \frac{x}{2}\rangle
\end{equation}
equivalent to~\eqref{eq:12} one finally obtains the desired
explicit expression for~\eqref{eq:11}:
\begin{multline}
   \label{eq:30}
   \langle q_1 | \rho_t | q_2 \rangle = \frac{1}{2\pi\hbar}\int dk\int
   dy\,e^{-(i/\hbar) ky}\exp \left\{{-\frac{\lambda\alpha^2}{2\hbar^2}k^2t}\right\}
   \\ \times
   \exp \left\{\frac{1}{8\alpha \Gamma_{t}^2} \left[ \left(
            (q_1-q_2)e^{-2\lambda\alpha
              t}-\frac{\Gamma_{t}}{2m\lambda\alpha}k \right)^2K_1 (t)+2
         (q_1-q_2) \left( (q_1-q_2)e^{-2\lambda\alpha
              t}-\frac{\Gamma_{t}}{2m\lambda\alpha}k \right)K_2 (t)+
         (q_1-q_2)^2K_3 (t) \right] \right\}
   \\ \times
   \left\langle {
        y+\frac{q_1+q_2}{2}+\frac{q_1-q_2}{2}e^{-2\lambda\alpha
          t}+\frac{kt}{2m} {\left( 1-\frac{\Gamma_{t}}{2\lambda\alpha t}
          \right)} \left|\vphantom{{\left(
                  1-\frac{\Gamma_{t}}{2\lambda\alpha t}
          \right)}}\rho_t^{S} \vphantom{{\left( 1-\frac{\Gamma_{t}}{2m\lambda\alpha}
          \right)}}\right|
        y+\frac{q_1+q_2}{2}-\frac{q_1-q_2}{2}e^{-2\lambda\alpha
          t}-\frac{kt}{2m} {\left( 1-\frac{\Gamma_{t}}{2\lambda\alpha t}
          \right)} } \right\rangle .
\end{multline}
\end{widetext}

\end{document}